\documentclass[reprint,aps,prd]{revtex4-2}
\usepackage{amsmath}
\allowdisplaybreaks[4]
\usepackage{multirow}
\usepackage{amsfonts}
\usepackage{xcolor}
\usepackage{comment}
\usepackage{graphicx}
\usepackage{amsmath}
\usepackage{siunitx}
\usepackage{lineno}
\usepackage{float}
\usepackage{ulem}
\usepackage[
    pdfauthor={Teng Zhang},
    pdftitle={7kHz},
    colorlinks=true,
    linkcolor=black,
    urlcolor=blue,
    citecolor=blue
]{hyperref}

\DeclareSIUnit{\sqrthz}{\ensuremath{\sqrt{\text{\hertz}}}}

\begin{document}

\renewcommand\texteuro{FIXME}

\allowdisplaybreaks[4]
\title{Towards observing the neutron star collapse with gravitational wave detectors}

\author{Teng Zhang}
\email{tzhang@star.sr.bham.ac.uk}
\affiliation{Institute for Gravitational Wave Astronomy, School of Physics and
Astronomy, University of Birmingham, Birmingham B15 2TT, United Kingdom}
\author{Ji\v{r}\'{i} Smetana}
\affiliation{Institute for Gravitational Wave Astronomy, School of Physics and
Astronomy, University of Birmingham, Birmingham B15 2TT, United Kingdom}
\author{Yikang Chen}
\affiliation{Gravitational Wave and Cosmology Laboratory, Department of Astronomy, Beijing Normal University, Beijing 100875, China}
\author{Joe Bentley}
\affiliation{Institute for Gravitational Wave Astronomy, School of Physics and
Astronomy, University of Birmingham, Birmingham B15 2TT, United Kingdom}
\author{William E. East}
\affiliation{Perimeter Institute for Theoretical Physics, Waterloo, Ontario N2L 2Y5, Canada}
\author{Denis Martynov}
\email{dmartynov@star.sr.bham.ac.uk}
\affiliation{Institute for Gravitational Wave Astronomy, School of Physics and
Astronomy, University of Birmingham, Birmingham B15 2TT, United Kingdom}
\author{Haixing Miao}
\email{haixing@star.sr.bham.ac.uk}
\affiliation{Institute for Gravitational Wave Astronomy, School of Physics and
Astronomy, University of Birmingham, Birmingham B15 2TT, United Kingdom}
\author{Huan Yang}
\email{hyang10@uoguelph.ca}
\affiliation{Perimeter Institute for Theoretical Physics, Waterloo, Ontario N2L 2Y5, Canada}
\affiliation{University of Guelph, Guelph, Ontario N2L 3G1, Canada}

\begin{abstract}
Gravitational waves from binary neutron star inspirals have been detected along with
the electromagnetic transients coming from the aftermath of the merger in GW170817. However, much is still unknown about the post-merger dynamics that connects these two sets of observables. This includes if, and
when, the post-merger remnant star collapses to a black hole, and what are the necessary conditions to power a short gamma-ray burst, and other observed electromagnetic counterparts. Observing the collapse of the post-merger neutron star  would shed led on these questions, constraining models for the short gamma-ray burst engine and the hot neutron star equation of state. In this work, we explore the scope of using gravitational wave detectors to measure the timing of the collapse either indirectly, by establishing the shut-off of the post-merger gravitational emission, or---more challengingly---directly, by detecting the collapse signal. For the indirect approach, we consider a kilohertz high-frequency detector design that utilises a previously studied coupled arm cavity and signal recycling cavity resonance. This design would give a signal-to-noise ratio of 0.5\,-\,8.6 (depending on the variation of waveform parameters) for a collapse gravitational wave signal occurring at 10\,ms post-merger of a binary at 50\,Mpc and with total mass $2.7 M_\odot$. This detector design is limited by quantum shot noise and the signal-to-noise ratio largely depends on the detector power, which is adopted as 4\,MW in this work. For the direct approach, we propose a narrow-band detector design, utilising the sensitivity around the frequency of the arm cavity free spectral range. To attain the maximal achievable quantum sensitivity, which is fundamentally limited by optical loss, we suggest the application of an optomechanical filter cavity that coverts the signal recycling cavity into a signal amplifier. The proposed detector achieves a signal-to-noise ratio of 0.3\,-\,1.9, independent of the collapse time. This detector is limited by both the fundamental classical and quantum noise with the arm cavity power chosen as 10\,MW.

\end{abstract}
\maketitle

\section{Introduction}
On August 17, 2017, the LIGO and Virgo detectors observed a gravitational wave signal, GW170817, from a binary neutron star system\,\cite{PhysRevLett.119.161101}.
Around 1.7 seconds after the coalescence, the Fermi Gamma-ray Burst Monitor observed a gamma-ray burst, GRB170817A\,\cite{170817A}, and initiated electromagnetic follow-up observations\,\cite{Abbott_2017}. 
This unprecedented observational campaign motivated a series of studies, exploring the implications for astrophysics, dense nuclear matter, gravitation, and cosmology\,\cite{PhysRevLett.119.161101}. In particular, GW170817 has been used as a ‘standard siren' to measure the Hubble constant\,\cite{NatureHC}, and to constrain the tidal deformation of neutron stars\,\cite{GWHC,Holz_2005,PhysRevD.74.063006,Nissanke_2010,nissanke2013determining}.

\begin{figure}[t]
\centering
  \includegraphics[width=1\columnwidth]{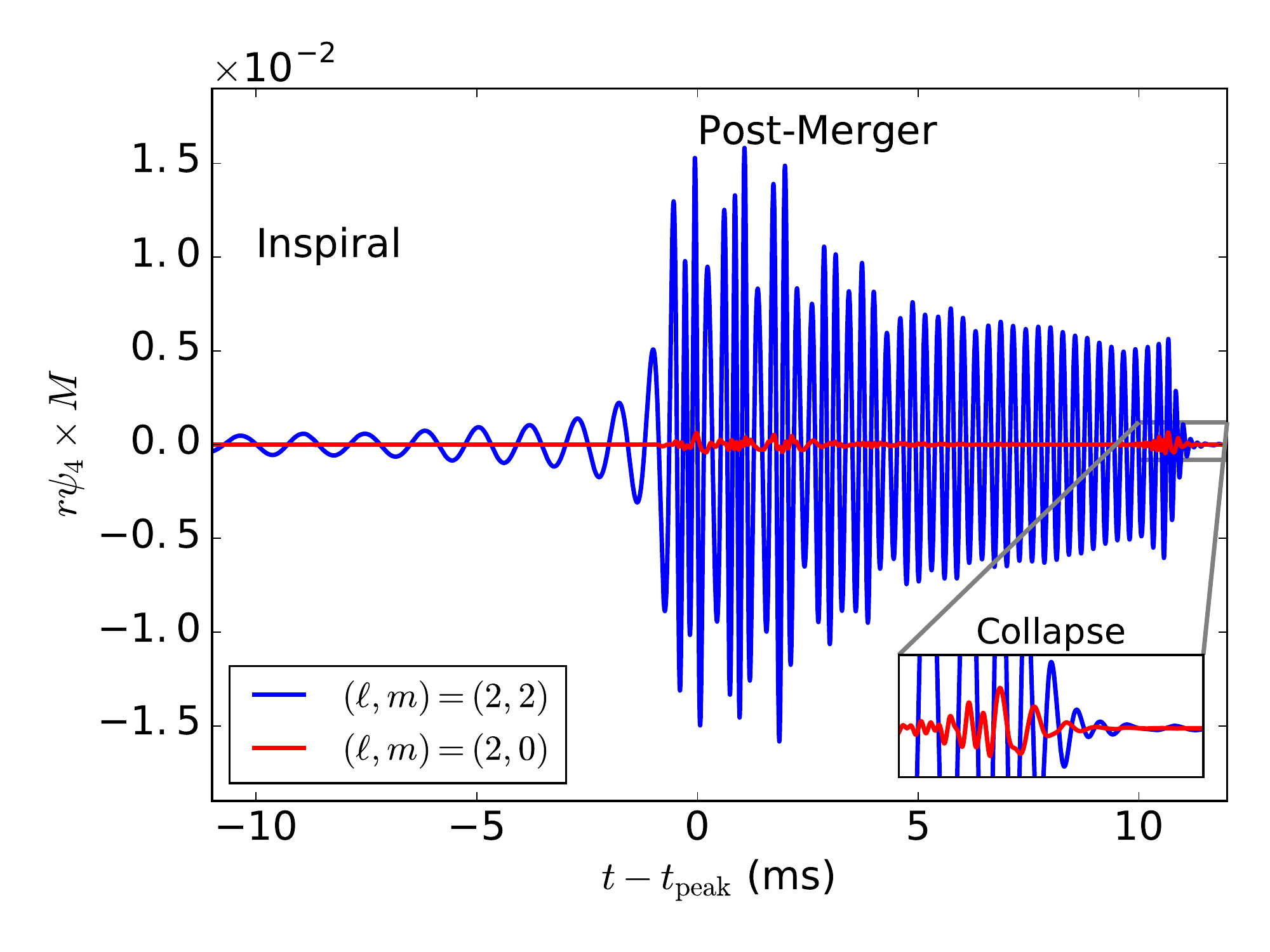}
  \includegraphics[width=1\columnwidth]{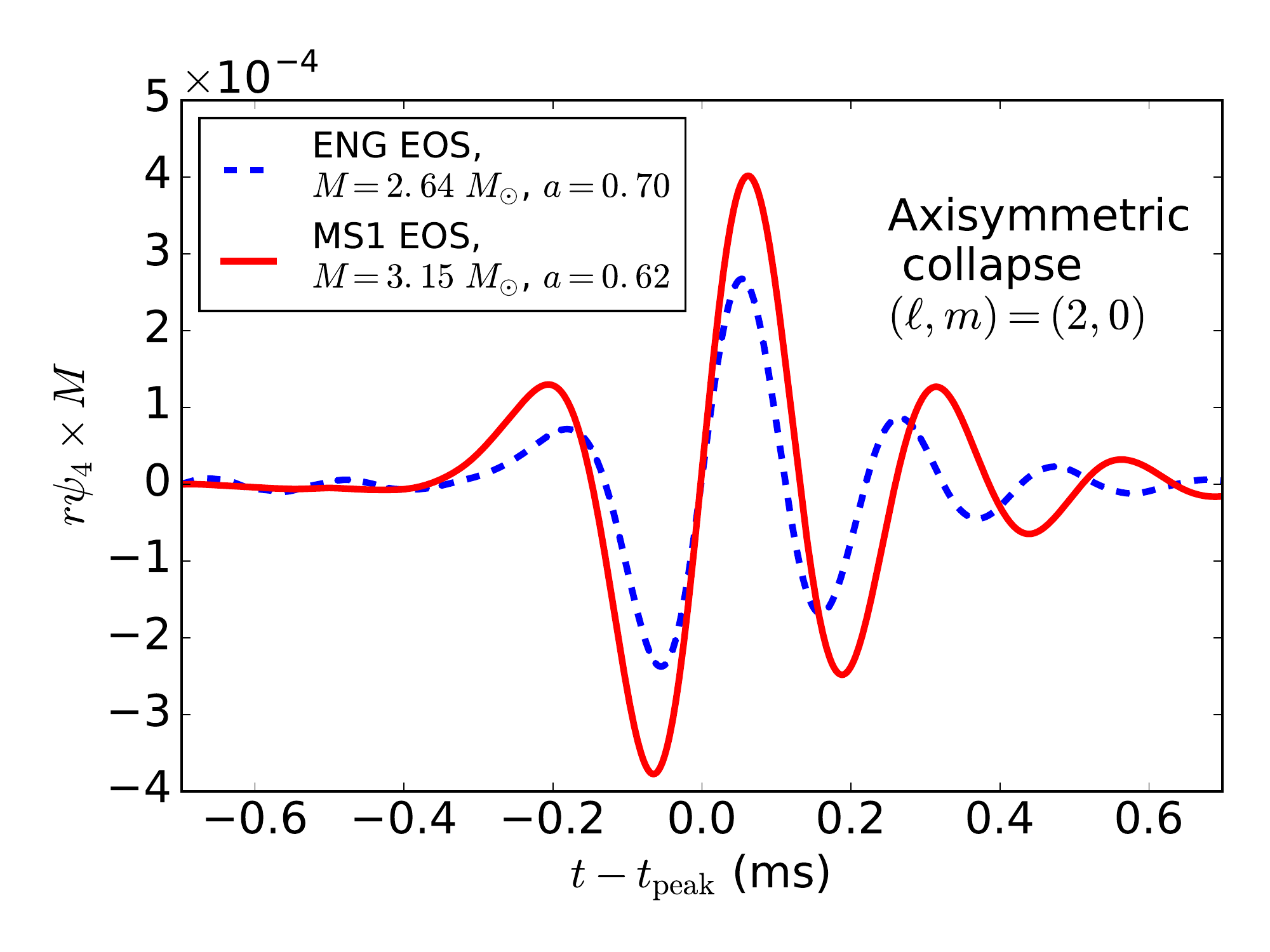}
    \caption{
        Top: An example binary neutron star merger waveform from
        Ref.~\cite{east2019binary} (in particular the Newman-Penrose scalar
        $\psi_4$ from the simulation with the HB EOS). In this case the
        post-merger remnant collapses to a black hole after $\sim10$ ms.  There
        is an $\ell=2$, $m=0$ gravitational wave component associated with the
        collapse (inset), though at this point it is still very small compared
        to the $\ell=2$, $m=2$ component which dominates throughout.
        Bottom: The $\ell=2$, $m=0$ component of the waveform from the collapse
        of unstable, uniformly rotating neutron stars. We show two examples with
        the ``ENG" and ``MS1" EOSs and different star masses. In the latter case, the
        $m=0$ component is about two-thirds the amplitude found in the top panel.
        We also find that the pre- and post-peak portions of the collapse
        waveform contribute roughly equally to the total power, and have
        similar frequencies.
    }
\label{fig:wav}
\end{figure}

Despite these groundbreaking multi-messenger observations from GW170817, much is still unknown about what happened in the 1.7 seconds between the merger time when the gravitational wave frequency became too high to be detectable by
Advanced LIGO/Advanced Virgo, and the occurrence time of  the gamma-ray burst, followed by a whole set of electromagnetic transients from the X-ray to the radio band. Generally speaking, a post-merger remnant star may eventually collapses to a black hole and remains  as a massive neutron star. Its fate and the timing of the possible collapse depends on many factors, including the neutron star equation of state (EOS) which is largely unknown; the dissipation and redistribution of angular momentum
through gravitational radiation, magnetic field instabilities, and viscous effects; as well as how much thermal support is created in the violent merger, and how quickly the remnant cools due to neutrino emission, \textit{etc.}\,\cite{Radice:2020ddv}.  Many of these factors depend on unknown microphysics, as well as complicated, nonlinear, and multiscale dynamics, which are rather preliminary in theoretical exploration. However, what happens in this stage is essential for generating the subsequent electromagnetic transients, so that its description is connected to the answers of many important open problems. For example, even though the multi-messenger observations of GW170817 have provided strong evidence that (at least some) short gamma-ray bursts come from binary neutron star mergers, the underlying burst engine and mechanism are still unknown~\cite{Paschalidis:2016agf}.
Though a leading explanation is that an accretion powered jet is launched by a black hole+ disk remnant, there are other proposals that the gamma-ray burst is launched by a magnetar\,\cite{Metzger:2007cd,bucciantini2012short,Rowlinson:2013ue}. Furthermore, if the post-merger remnant does collapse to a black hole, its lifetime also affects how massive the accretion disk is, how long it is irradiated by neutrino emission from the post-merger star, and the evolution of the magnetic fields.
These factors  all impact the subsequent electromagnetic signals, including the synthesis of r-process material and the associated kilonovae transient~\cite{Metzger:2016pju,Siegel:2019mlp}.
On the other hand, assuming the collapse does happen, 
a measurement of its timing may be combined with the electromagnetic observations to constrain
existing electromagnetic transient models, or to motivate new ones. The timing information can also be used
to constrain the neutron star EOS\,\cite{sarin2020gravitational}, especially in the regime of higher densities and temperatures than in the ones in the inpiral stage. Therefore, it is astrophysically interesting to explore the design of a detector that is capable of identifying the prompt or delayed collapse of binary neutron star merger remnant. As we shall see, this task is particularly challenging, as the frequencies of the collapse gravitational wave signals are
generally higher than the main post-merger signals, and the amplitudes are weaker. 

We propose two different methods of identifying or constraining the timing of
the collapse. In the indirect approach, we assume that we already have 
an accurate estimate for the post-merger waveform of
hypermassive/supramassive neutron stars, so that we may infer the post-merger
collapse from a non-detection of the post-merger signal with sufficient sensitivity.  In the direct approach, we search for the collapse signal within
the post-merger data using the collapse waveform, with a successful detection naturally giving rise to the
timing information. We  discuss
the performance of both methods with the detector designs explored in
this work. The detector design for indirect approach follows the conventional idea in Ref.\,\cite{PhysRevD.99.102004,ackley2020neutron}, which takes the coupled arm cavity and signal recycling cavity at 2\,-\,3\,kHz.
For the direct approach, the detector is a narrow band detector targeting on exploring the sensitivity around the frequency of arm cavity free spectral range (FSR). It is largely different from the current existing and future planned broadband gravitational wave detectors,\textit{e.g.} Advanced LIGO\,\cite{LIGO2015}, Advanced Virgo\,\cite{Acernese_2014}, KAGRA\,\cite{Akutsu:2019aa}, Einstein Telescope\,\cite{Hild_2011,ET-D,ET2020} and Cosmic Explorer\,\cite{reitze2019cosmic}. 
This is the first attempt of the detector design around the FSR.

The paper is organised as follows: in Sec.~\ref{sec:WF}, we describe the
typical gravitational wave waveform from the collapse of a massive neutron star. In
Sec.~\ref{Sec:detection}, we describe two detection strategies and summarise
the resulting signal-to-noise rations with the two optimal gravitational
wave detectors proposed in this work; In Sec.~\ref{sec:QuantumLoss}, we study
the quantum noise that is bounded below by quantum losses. For the direct approach, we introduce a scheme that can saturate the quantum loss limit
at the FSR. In Sec.~\ref{sec:classical}, we discuss the classical noises in the two
detectors.

\section{Collapse Waveform}
\label{sec:WF}
\def\we#1{{\textcolor{cyan}{\bf WE: #1}}}

Depending on its total mass, a binary neutron star system may promptly collapse
to a black hole after merger, or form a long-lived post-merger remnant that
 only collapses after redistributing/losing angular momentum to
gravitational wave emission and magneto-vicious effects and/or radiative
cooling. A third possibility is that the neutron star EOS can support a
stable remnant.
At the onset of the collapse, the remnant star may still have an oscillating
quadrupole moment due to its binary origin, so that the collapse process is not
fully axisymmetric.  As a result, we expect both the $\ell=2,\ |m|=2$ (or
``22") and $\ell=2,\
m=0$ (or ``20") spherical harmonics to contribute to the gravitational wave
emission.  The 22 part of the waveform mainly describes the continuous
evolution and damping of the remnant quadrupole moment during the collapse
process, and the 20 component comes from the collapsing, angular-averaged
remnant star, which is non-spherical due to its spin. Right after the binary
neutron star collision, the quadrupole moment emission of the remnant should
play a significant role in the collapse waveform; at late times the remnant
quadrupole moments damp out, in which case the collapse waveform should
be dominated by the 20 piece.
This is illustrated in the top panel of Fig.~\ref{fig:wav}, where we show these
two contributions for an sample binary neutron star merger simulation from
Ref.~\cite{east2019binary}. In this case, the collapse happens roughly 10\,ms
after merger, and the 22 mode contribution is still much larger than that of
the 20.  The former has been well studied in numerous cases using binary
neutron star simulations \cite{baiotti2017binary}, while the latter
has been less well-studied since it is typically small. In order to
characterize the lower bound on the expected collapse gravitational wave
signal---occurring when the remnant is able to completely reach rigid rotation
and cool prior to forming a black hole---we construct several unstable
equilibrium solutions describing uniformly rotating neutron stars, and simulate
their collapse. We show the resulting gravitational waveforms for two examples
with masses and spins chosen to match that expected from a binary merger and
different EOSs in the bottom panel of Fig.~\ref{fig:wav}.  These simulations
are carried out with the code described in Ref.~\cite{East:2011aa}, using the
same methods as in Ref.~\cite{east2019fate} (in particular, the equilibrium
star solutions are constructed using the RNS code~\cite{Stergioulas:1994ea}).
We find that the amplitude of the 20 mode is not significantly smaller (within
a factor of 2) from the merger case, and is characterized by the black hole
quasinormal mode ringing, with comparable contribution from the pre- and
post-peak power. The gravitational waves are also similar (after normalizing
by total mass) to those found in Refs.~\cite{east2019fate,Baiotti:2007np}, which
considered the (in some cases induced) collapse of lower mass rotating neutron
stars with various EOSs, indicating these results are relatively insensitive to
the particular EOS, etc. We will use this to guide the waveform
model described below.
 
We start by describing the post-merger waveform model mathematically. Despite the uncertainties in the neutron star's EOS and the complicated magneto-hydrodynamical processes governing the post-merger dynamics, it is generally accepted that the peak mode is one of the most robust features in the post-merger waveform\,\cite{bauswein2012measuring,hotokezaka2013remnant,lehner2016unequal}. This peak mode is associated with the emission from the rotating quadrupole moment of the remnant star, which usually dominates the waveform spectra after the first several milliseconds following the collision.

We phenomenologically model the peak mode signal as a decaying sinusoid for a source at distance $d$:
\begin{align}
h = A \frac{50 {\rm Mpc}}{d} \sin (2\pi f t) e^{-\pi f t/Q} e^{-\pi f \tau/Q}\,,
\end{align}
where the amplitude $A$, mode frequency $f$, and quality factor $Q$ all depend on the star mass and EOS,
and we take the convention that the waveform amplitude is specified for an observer on axis. 
Here, the delay time $\tau$ is the time separation between the star merger and the collapse, i.e., the collapse time. This 
is the part of the post-merger waveform that we would have detected if there were no collapse happening. If we pick a typical EOS ``TM1" and choose an equal-mass binary with total mass $M=2.7 M_\odot$, the parameters take the values
\begin{align}\label{eq:peak}
A \approx 2.5\times 10^{-22}, \quad f \approx 2.8 \,{\rm kHz}, \quad Q \approx 34\,.
\end{align}
We shall use this set of values as ``canonical" values for later demonstration. Values for more EOSs can be found in Ref.\,\cite{yang2018gravitational}. For example, the amplitude changes by about a factor of two for the 5 EOSs considered in Ref.\,\cite{yang2018gravitational}, ranging from the stiff side to the soft side of the EOS spectrum.

The collapse waveform contains two parts: the 22 mode and 20 mode. As the post-collapse waveform is very close to black hole ringdown, the 22 and 20 mode frequencies and damping rates are well approximated by black hole quasinormal mode frequencies. For a black hole with mass $\sim 2.7 M_\odot$ and dimensionless spin $a \sim 0.7$ (typical for a post-collapse black hole), the corresponding mode frequencies and quality factors are
\begin{align}\label{eq:2022}
& f_{22} \sim 6.4 \,{\rm kHz}, \quad f_{20} \sim 4.7\, {\rm kHz} \nonumber \\
& Q_{22} \sim 3.2, \quad Q_{20} \sim 2.5\,.
\end{align}
If we allow a range of spins $0.65 -0.80$ for prompt \cite{kiuchi2009long,rezzolla2010accurate,bernuzzi2014mergers,kastaun2013black} and delayed collapses \cite{sekiguchi2016dynamical}, and $\sim 20\%$ variation between the mass thresholds of prompt and uniform-rotation collapses \cite{bauswein2013prompt,rezzolla2018using}, the variation in the quasinormal frequency is approximately $30\%$, and the variation of quality factor is $\le 20\%$ \cite{yang2012quasinormal}.  

Based on the discussion above,
we shall take as a representative value for the amplitude of the 20 mode $A_{20} \sim 0.8 \times 10^{-23}$, though it is reasonable to expect at least a factor of 2 variation depending on the specific EOS and star mass.
On the other hand, if we make the observation that the amplitude of $\psi_4$ for the 22 mode is roughly continuous before and after the collapse \cite{baiotti2008accurate}, we can estimate $A_{22}$ as $A_{22} \sim A (f/f_{22})^2$. Notice that this expression may underestimate the 22 mode amplitude if the collapse happens in the early post-merger phase ($\le 20$ms), where the peak mode only constitutes a sub-dominant fraction of the post-merger spectrum \footnote{The signal-to-noise ratio of the peak mode is roughly $1/2$-$1/3$ of the signal-to-noise ratio of the whole post-merger waveform \cite{PhysRevD.99.102004}}. Other short-lived modes may also deform the quadrupole moments and affect the post-collapse 22 mode. 

In summary, the post-collapse 22 waveform is 
\begin{align}
h_{22} = A_{22} \frac{50 {\rm Mpc}}{d} \sin (2\pi f_{22} t) e^{-\pi f_{22} t/Q_{22}} e^{-\pi f \tau/Q}\,.
\end{align}
The collapse 20 waveform is
\begin{align}
h_{20} = A_{20} \frac{50 {\rm Mpc}}{d} \sin (2\pi f_{20} |t|) e^{-\pi f_{20} |t|/Q_{20}} \,.
\end{align}
We note that the axisymmetric part of the star is not sensitive to the collapse time, so that in this model, $h_{20}$ is assumed to be independent of $\tau$.
\begin{figure}[t]
\centering
  \includegraphics[width=1\columnwidth]{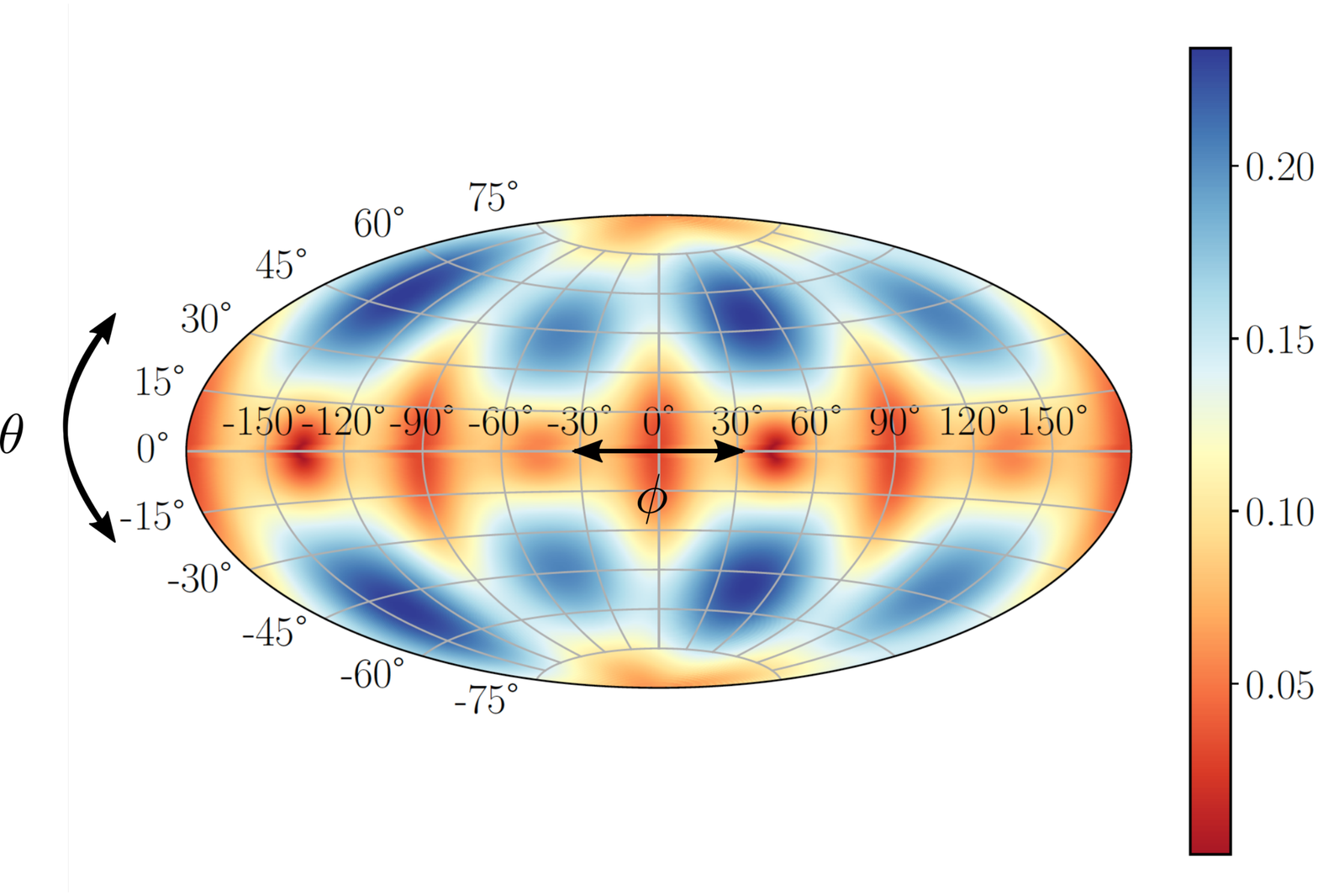}
\caption{The square root of the antenna power pattern of a gravitational wave detector for signals from sky location described by $\theta$ and $\phi$ at frequency of the arm cavity FSR.}
\label{fig:AntFSR}
\end{figure}

\section{Detection strategies}\label{Sec:detection}

In order to determine the collapse time, there are at least two detection strategies based on the measurement of gravitational waves. 
\begin{figure*}[t]
\centering
  \includegraphics[width=2.0\columnwidth]{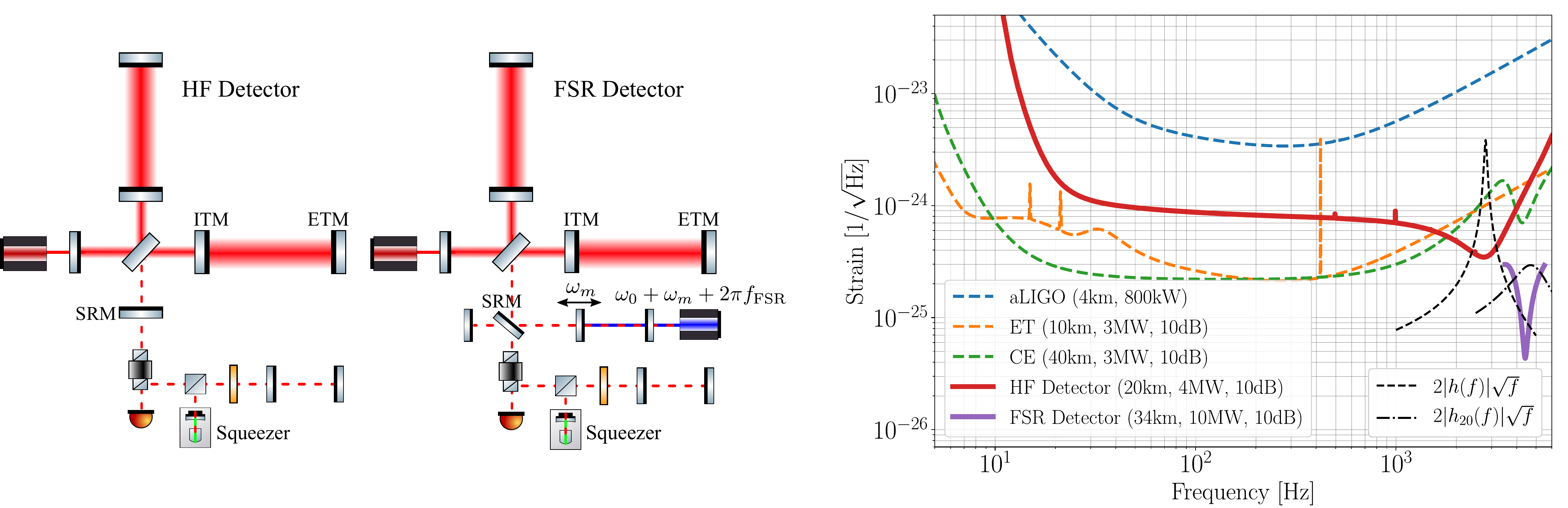}
\caption{The left part of the figure shows two gravitational wave detector configurations, the 20\,km HF detector and the 34\,km FSR detector, which are optimal for detecting the peak mode of the post-merger waveform and the 20 mode of the collapse waveform. The right part of the figure shows the strain sensitivity of the two configurations in comparison with Advanced LIGO, Einstein Telescope and Cosmic Explorer. 
The strain sensitivity of the FSR detector is for sources at $\theta=45^{\circ}, \phi=45^{\circ}$, where the antenna response is at a maximum.
The arm cavity powers of the HF and FSR detectors are 4\,MW and 10\,MW, respectively. The assumed observed squeezing levels are 10\,dB. The other detector parameters are shown in Table~\ref{ta:par}. The amplitudes of the peak mode and 20 mode with parameters listed in Eq.~\eqref{eq:peak} and Eq.~\eqref{eq:2022} are shown as dashed and dot-dashed curves, respectively. }
\label{fig:detectorScheme}
\end{figure*}

The first method (``indirect" approach), assumes that we can accurately predict the post-merger waveform based on the inspiral measurement. Such an assumption relies on future advances in theoretical modelling of the post-merger processes and gravitational-wave measurements of the star EOS. If this assumption is satisfied, starting from any delay time $\tau$, we can compare the inferred waveform signature with the actual waveform. If with enough confidence we no longer detect the inferred emission from the hypermassive neutron star starting from time $\tau$, the remnant star must have collapsed.

The second method (``direct" approach) is to directly identify the signal of the collapse waveform in the entire post-merger process, either the 20 component or the 22 component, or both. The threshold signal-to-noise ratio (SNR) depends on the time width of the search window  and the requirement on the false alarm rate.
As the duration of the collapse waveform is on the order of milliseconds, successful detection with the direct method will be able to pin down the collapse time to $\mathcal{O}(\rm ms)$ accuracy.

For both strategies, we calculate the achievable SNR of a detector for the three types of waveform mentioned earlier: the post-merger peak mode, the collapse 20 mode and the collapse 22 mode. The detector strain sensitivity to a particular source in the sky relies on the antenna response, $F_{+/\times}$, which is a function of source location and gravitational wave signal frequency. A detailed calculation of the antenna response can be found in Ref.\,\cite{PhysRevD.96.084004}. When the gravitational wave signal frequency is much smaller than the arm cavity FSR, the long wavelength approximation is applicable, under which the square root of the antenna power pattern, $\sqrt{|F_{+}|^2+|F_{\times}|^2}$, is maximal and equals to 1 for gravitational waves at normal incidence, \textit{i.e.} the polar angle $\theta=0$. Considering a long baseline facility, important features of the antenna response appear around the frequency of the arm cavity FSR. Fig.~\ref{fig:AntFSR} shows the square root of the antenna power pattern, $\sqrt{|F_{+}|^2+|F_{\times}|^2}$ at frequency of the arm cavity FSR. Here, $\sqrt{|F_{+}|^2+|F_{\times}|^2}$ is 0 for $\theta=0$, while the maximal response is around 0.25 for a signal source at $\theta=45^{\circ}, \phi=45^{\circ}$. The maximal SNR is characterised as 
\begin{equation}
{\rm SNR}=2\sqrt{\int \frac{|h(f)|^2|(|F_{+}|^2+|F_{\times}|^2)|_{\rm max}}{S_{hh}}df}\,,
\end{equation}
where $S_{hh}$ is the signal-referred power spectral density for the noise. In the following discussion, the SNR is calculated for the binary sources at 50\,Mpc and with total mass $2.7 M_\odot$.  The "canonical" parameters are listed in Eq.~\eqref{eq:peak} and Eq.~\eqref{eq:2022}.  

\begin{figure*}[t]
\centering
  \includegraphics[width=2\columnwidth]{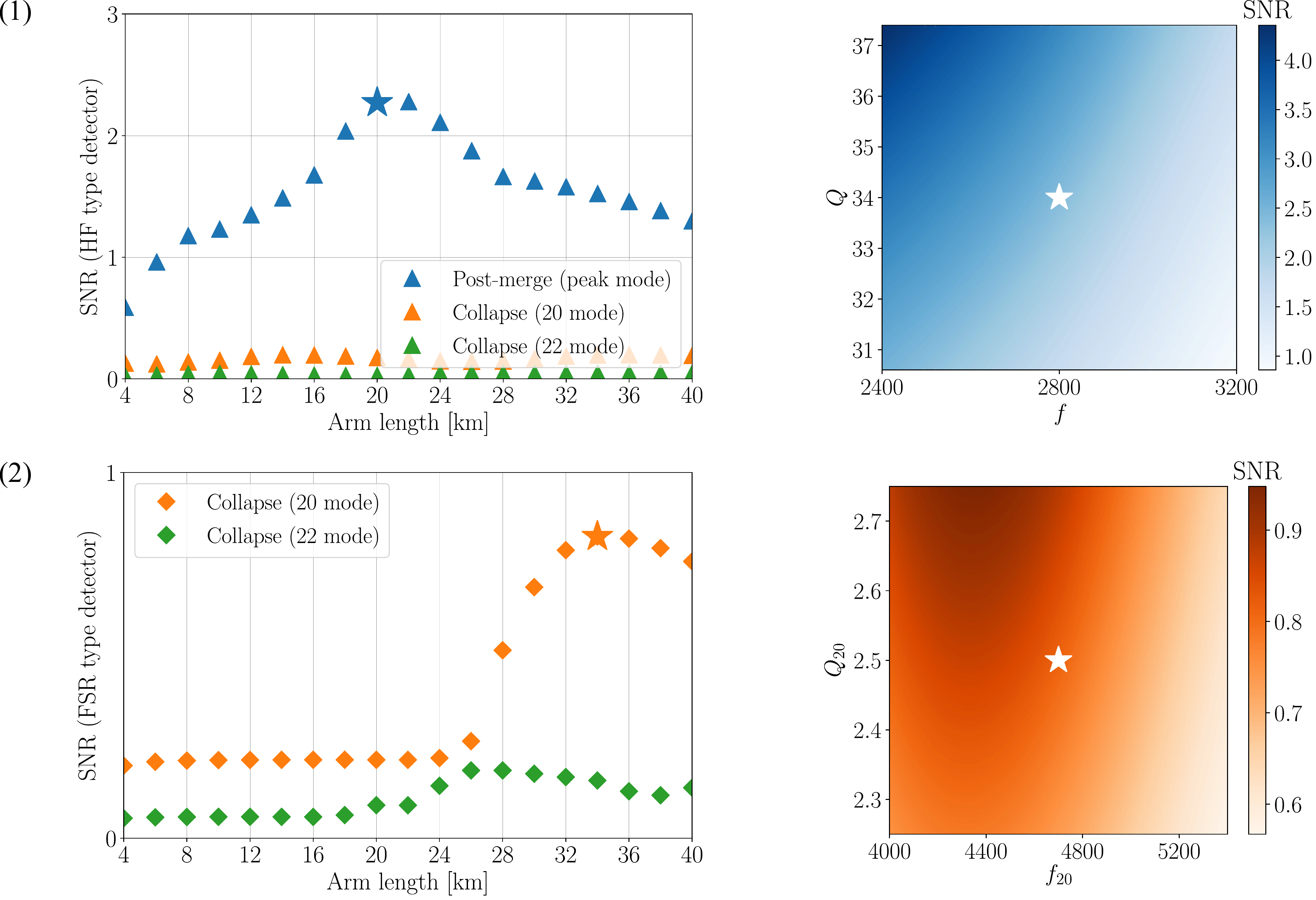}
\caption{(1) Left: The SNR of the HF type detector for the peak mode of the post-merger waveform, and the 20 and 22 modes of the collapse waveform as a function of arm length. The waveform parameters are given as the “canonical” values in Eq.~\eqref{eq:peak} and Eq.~\eqref{eq:2022} and a delay time of 10\,ms. The maximal SNR (2.3) for the peak mode is at around 20\,km marked with the star, which is chosen for the arm length of the HF detector. The other detector parameters are listed in Table.~\ref{ta:par}; Right: With the HF detector, the SNR for the peak mode of the post-merge waveform which has a 30\% variation in the qausinormal frequency $f$ and a 20\% variation in the quality factor. The white star corresponds to the star in the left part.
(2) Left: The SNR of the FSR type detector for the 20 and 22 modes of the collapse waveform as a function of arm length.  The maximal SNR (0.8) for the 20 mode is at around 34\,km marked with the star, which is chosen for the arm length of the FSR detector. The waveform parameters are given as the “canonical” values in Eq.~\eqref{eq:2022} and a delay time of 10\,ms. The other detector parameters are listed in Table.~\ref{ta:par}; Right: With the FSR detector, the SNR of the 20 mode of the collapse waveform which has a 30\% variation in the qausinormal frequency $f_{20}$ and a 20\% variation in the quality factor. The white star corresponds to the star in the left part.}
\label{fig:SNRfq}
\end{figure*}
\begin{figure}[t]
\centering
  \includegraphics[width=1\columnwidth]{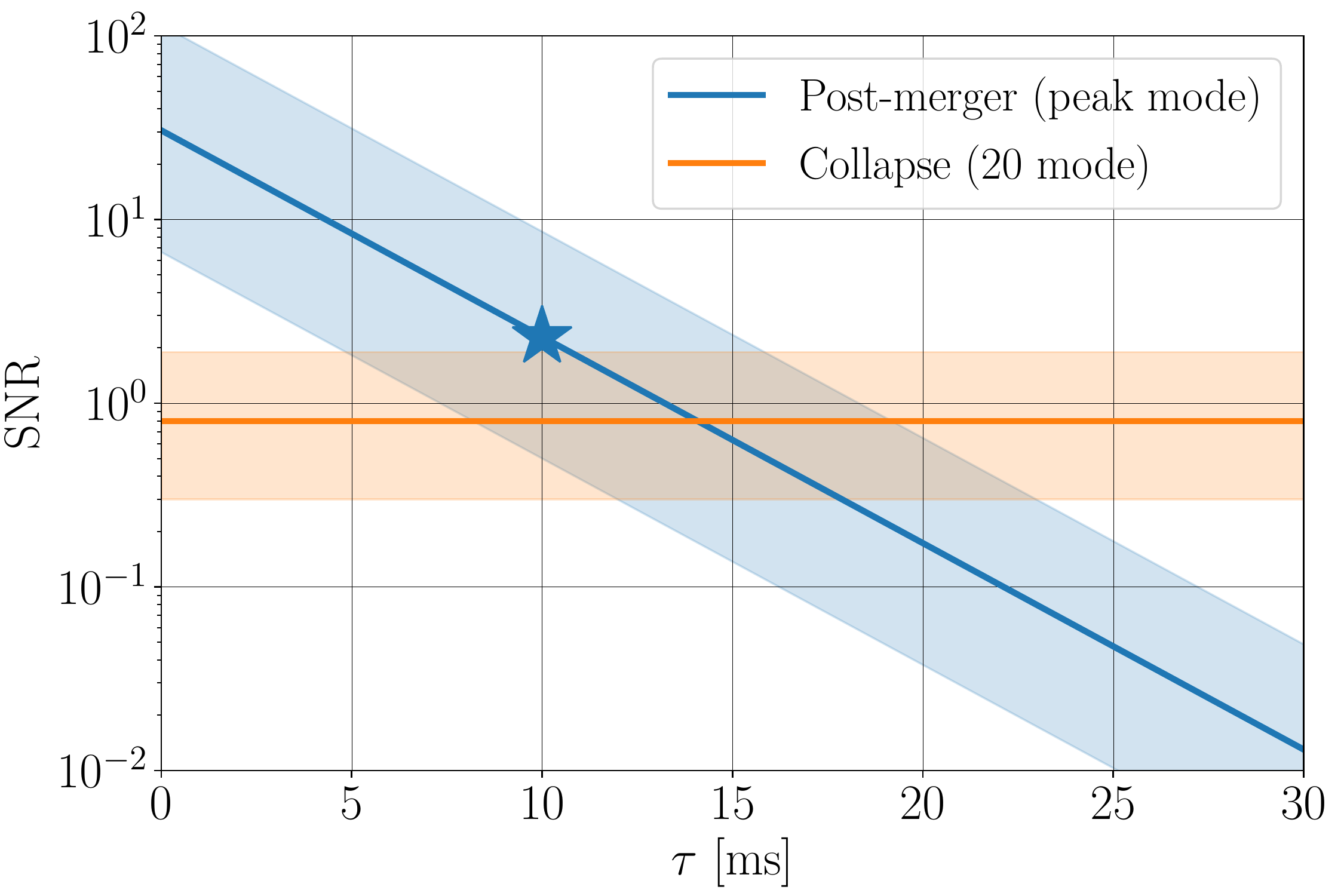}
\caption{ The SNR of the HF detector to peak mode of the post-merge waveform and FSR detector to the 20 mode of the collapse waveform as a function of delay time. The solid lines are for waveform parameters given in Eq.~\eqref{eq:peak} and Eq.~\eqref{eq:2022}. The shaded areas give boundaries of minimal and maximal SNR taking into account the variation in the qausinormal frequency, the quality factor and a factor of 2 variation of the amplitudes of waveforms. At $\tau=10$\,ms, the SNR for peak mode is in the range of 0.5\,-\,8.6, for 20 mode is in the range of 0.3\,-\,1.9.}
\label{fig:SNR}
\end{figure}

In Fig.~\ref{fig:detectorScheme}, we give two detector designs, for the indirect and direct approaches respectively. For the former, the high-frequency (HF) detector design follows the one proposed in Ref.\,\cite{PhysRevD.99.102004}, which assumes the current existing technologies but with 4\,MW arm power and 10\,dB observed squeezing. 
As it turns out, the optimal length for measuring the peak mode of the post merger waveform is around 20\,km, which gives SNR of 2.3 for the peak mode with $\tau=10$\,ms as shown in Fig.~\ref{fig:SNRfq}. This is consistent with the finding in Ref.\,\cite{PhysRevD.99.102004}, depending on the variation in the quasinormal frequency within 30\% and variation in the quality factor within 20\%. The resulting SNR can vary from 1.0 to 4.3. And allowing a factor of 2 variation of waveform amplitude, SNR can vary from 0.5 to 8.6. However, the SNR for 20 or 22 mode of collapse waveform from this design are far from allowing a direct detection of the collapse. 

For the direct approach, an alternative design making use of the detector sensitivity at the first FSR turns out to be a better option. For such an FSR detector design, to achieve an SNR $\sim$ 1, we have to assume 10\,MW arm power and 10\,dB observed squeezing. We also implement a new quantum technique based upon the active optomechanical filter, which allows us to achieve a loss-limited quantum sensitivity around the FSR. The classical noises are estimated based on proposed technology for future gravitational wave detectors, \textit{e.g.} Voyager\,\cite{Adhikari_2020}, Einstein Telescope\,\cite{Hild_2011,ET-D,ET2020} and Cosmic Explorer\,\cite{reitze2019cosmic}.  As shown in Fig.~\ref{fig:SNRfq}, the optimal length is 34\,km, giving an SNR of 0.8 for the 20 mode of the canonical collapse waveform. We also show that  by considering the variation in the quasinormal frequency and the quality factor, within 30\% and 20\% respectively, the SNR vary from around 0.6 to around 0.95. Taking into account a factor of 2 variation of the waveform amplitude, the FSR detector gives constant SNR for the 20 mode in the range of 0.3\,-\,1.9. The SNR for the 22 mode with $\tau=10$\,ms is small, unless a short collapse time is expected. In fact, the SNR of measuring the peak mode of the post-merge waveform with the indirect appraoch is expect to increase with a shorter collapse time. As shown in Fig.~\ref{fig:SNR}, in which the shaded areas indicate the boundary of the SNR taking into account variation in quasinormal frequency, quality factor and amplitude, the HF detector is preferred for the indirect approach when $\tau < 16$\,ms. 

In the following sections, we will discuss the quantum noise and classical noise of the two detector designs.

\section{Quantum noise and quantum limit from optical losses at the FSR}\label{sec:QuantumLoss}

At high frequencies, the gravitational wave detector sensitivity is critically limited by quantum noise.
The quantum noise consists of quantum radiation pressure noise and quantum shot noise. The former, which results from the optomechanical interaction between the light and the mirror, dominates at low frequencies. Above the interferometer bandwidth, the quantum shot noise dominates. Given the arm cavity power, the quantum shot noise limited sensitivity can also be further tailored by choosing parameters of the signal recycling cavity (SRC), \textit{i.e.} the cavity length and finesse, which shapes the detector response to gravitational waves. By tuning the SRC parameters, we can optimise the quantum sensitivity for the figure of merit at each arm length. Further reduction of the quantum noise can be realised by using squeezed light.

However, the quantum noise cannot be infinitely suppressed but is limited by a lower bound from the optical losses, especially including arm cavity loss and SRC loss \cite{PhysRevX.9.011053}. Furthermore, the external output loss will limit the observed squeezing level. We assume 10\,dB of observed squeezing, which implies a maximum of 10\% output loss.

The arm cavity loss limit can be calculated as 
\begin{equation}
S_{hh}^{\rm arm}=\frac{\hbar c^2}{4L^2\omega_0 P_{\rm arm}}\epsilon_{\rm arm}\,,
\end{equation}
where $c$ is the speed of light, $\hbar$ is the Plank constant, $L$ is arm length, $P_{\rm arm}$ is the arm circulating power and $\epsilon_{\rm arm}$ quantifies the arm cavity loss. $\epsilon_{\rm arm}= 100$\,ppm is assumed in both detector designs.
The SRC loss limit can be modelled as the shot noise spectral density of the differential mode of the arm cavities scaled with $\epsilon_{\rm SRC}$,
\begin{equation}\label{eq:SRCloss}
S_{hh}^{\rm SRC}=\frac{\hbar c^2T_{\rm itm}}{16L^2\omega_0 P_{\rm arm}}\left[1+\frac{(\Omega-2n\pi f_{\rm FSR})^2}{\gamma^2}\right]\epsilon_{\rm SRC}\,,
\end{equation}
where $\epsilon_{\rm SRC}$ quantifies the SRC loss, $f_{\rm FSR}=\frac{c}{2L}$ is the FSR of the arm cavity, $n$ denotes the number of FSR away from the carrier frequency, and $\gamma=\frac{cT_{\rm itm}}{4L}$ is the arm cavity bandwidth. 

\begin{figure}[t]
\centering
  \includegraphics[width=1\columnwidth]{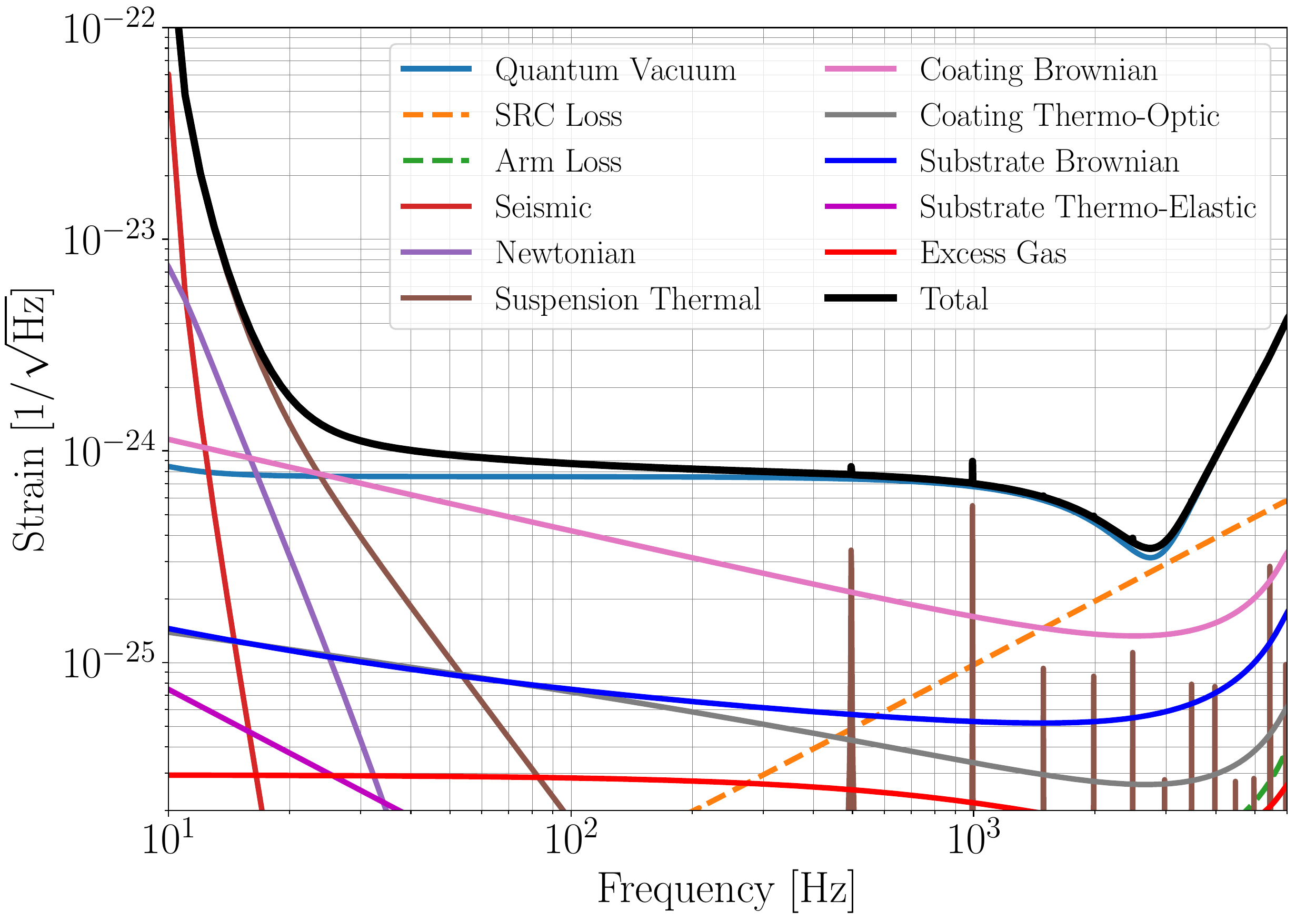}
\caption{Noise Budget of the HF detector. The detector strain sensitivity is for optimal orientated sources at $\theta=0^{\circ}, \phi=0^{\circ}$. The detector parameters are listed in Table.~\ref{ta:par}.}
\label{fig:HFbudget}
\end{figure}

As shown in Fig.~\ref{fig:HFbudget} (the noise budget of the HF detector), the high frequency quantum noise dip is limited by the SRC loss, which is assumed to be 1000\,ppm. The dip around 3\,kHz is shaped through the coupled cavity resonance between the arm cavity and the SRC in the same way as in Ref.\,\cite{PhysRevD.99.102004}. At the dip frequency, the sidebands resonate in the SRC, i.e. the accumulated phase reflecting from the arm cavity and traveling through a round trip in a long SRC equals to $2\pi$. The resonance frequency is approximately, $\frac{c\sqrt{T_{\rm itm}}}{2\sqrt{LL_{\rm SRC}}}$, where $L_{\rm SRC}$ is the SRC length. The depth of the dip is determined by the SRC finesse. The choices of parameters for the HF detector is shown in Table.~\ref{ta:par}. 

Towards higher frequencies, the SRC loss limit increases until $\Omega$ approaches the first FSR, where the noise returns to the level at zero frequency according to Eq.~\ref{eq:SRCloss}. To enable the best achievable sensitivity, an addition scheme which allows for saturating the loss limit is required. As shown in Ref.\,\cite{zhang2020broadband}, a proper arrangement of an optomechanical filter cavity allows the quantum noise to saturate the loss limit by enhancing the signals inside the SRC, without sacrificing the detector bandwidth defined by the arm cavity. The mechanism functions by cancelling the phase gained by the signal sidebands reflecting from the arm cavity through a negative dispersion from the optomechanical filter cavity. Therefore, the signal sidebands will resonate in the SRC at all frequencies and their amplitudes at the output are amplified by a constant gain, $1/\sqrt{T_{\rm SRM}}$ with $T_{\rm SRM}$ being the power transmissivity of the signal recycling mirror.

\begin{figure}[b]
\centering
  \includegraphics[width=0.9\columnwidth]{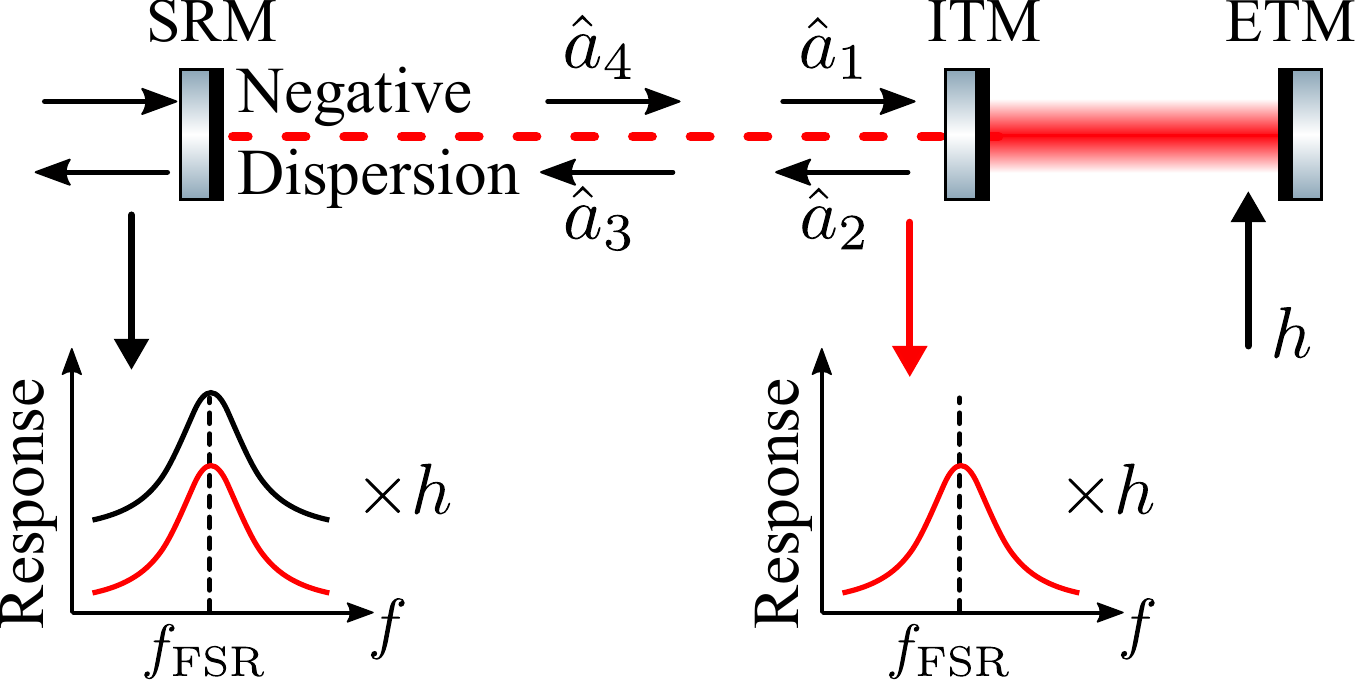}
\caption{The diagram shows the coupled arm cavity and SRC. The SRC can be converted into a signal amplifier around the arm cavity FSR frequency with frequency dependent negative dispersion inside for cancelling the accumulated phase of the sidebands reflected from the arm cavity. The amplification factor is $1/\sqrt{2 T_{\rm SRM}}$.}
\label{fig:SRCgain}
\end{figure}

\begin{figure}[t]
\centering
  \includegraphics[width=1\columnwidth]{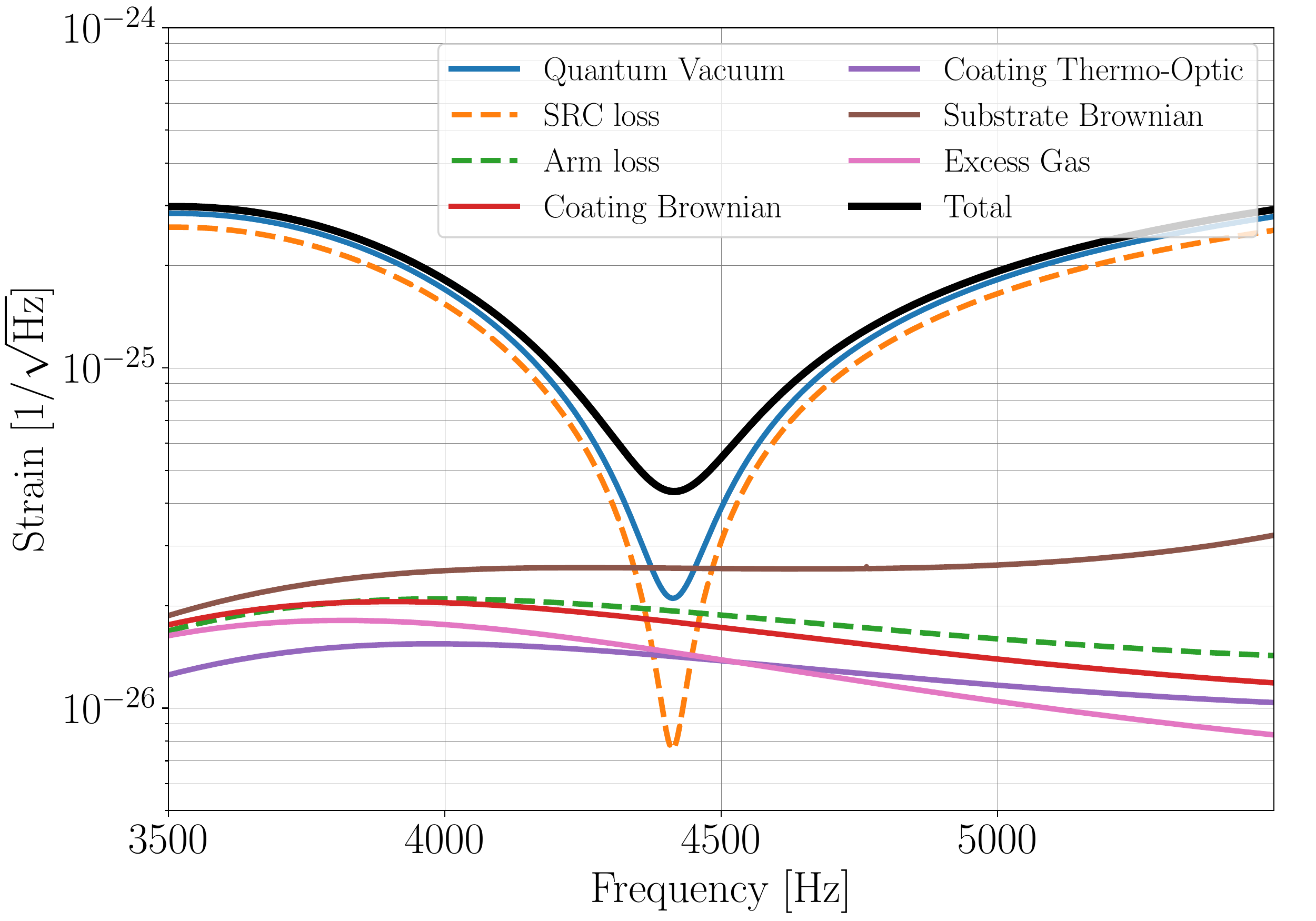}
\caption{Noise Budget of the FSR detector design. The detector strain sensitivity is for optimal orientated sources at $\theta=45^{\circ}, \phi=45^{\circ}$. The detector parameters are listed in Table.~\ref{ta:par}.}
\label{fig:FSRbudget}
\end{figure}

As we will illustrate, a scheme, similar to the one considered in Ref.\,\cite{zhang2020broadband}, can also work in our case for saturating the SRC loss limit around the FSR frequency. As shown in Fig.~\ref{fig:detectorScheme}, an optomechanical filter cavity is placed inside the signal recycling cavity. This filter cavity is pumped with a laser at frequency $\omega_0+\omega_m+2\pi f_{\rm FSR}$, where $\omega_m$ is the mechanical resonant frequency of the mechanical oscillator. Compared to Ref.\,\cite{zhang2020broadband}, the pump laser frequency is shifted by one FSR. 
When there is
\begin{equation}
\Omega-2\pi f_{\rm FSR}\ll \gamma_f \ll \omega_m \,,
\end{equation}
which is the so-called resolved-sideband regime, the input/output relation of the optomechanical filter for the sideband at $\Omega$ is
\begin{equation}\label{eq:optIO}
\hat{a}_{4}(\Omega)\approx-\frac{\gamma_{\rm opt}-i(\Omega-2\pi f_{\rm FSR})}{\gamma_{\rm opt}+i(\Omega-2\pi f_{\rm FSR})}\hat{a}_{3}(\Omega)\,,
\end{equation}
where $\gamma_{\rm opt}$ is the negative mechanical damping rate due to the optomechanical interaction.
The sidebands reflect from the arm cavity and there is
\begin{equation}\label{eq:armIO}
\hat{a}_{2}(\Omega)\approx \frac{\gamma+i(\Omega-2\pi f_{\rm FSR})}{\gamma-i(\Omega-2\pi f_{\rm FSR})}\hat{a}_{1}(\Omega)\,,
\end{equation}
When $\gamma_{\rm opt}=\gamma$, the phases from reflecting off the arm cavity and the optomechanical filter cavity cancel out. Therefore, the signal sideband at $\Omega$ becomes resonant, and the amplitude is amplified by $1/\sqrt{T_{\rm SRM}}$. However, the sideband at $-\Omega$ does not acquire the negative dispersion according to Eq.~\ref{eq:optIO}. Therefore, when measuring the combination of sidebands around $\pm f_{\rm FSR}$ using homodyne detection, we will get a factor of $\sqrt{2}$ degradation of the signal amplitude, similar to the detuned SRC case. 
The resulting quantum noise spectral density around the frequency of arm cavity FSR can be calculated as 
\begin{equation}
\begin{split}
S_{hh}^{\rm FSR}=\frac{\hbar c^2T_{\rm itm}T_{\rm SRM} e^{-2r}}{8L^2\omega_0 P_{\rm arm}}\left[1+\frac{(\Omega-2n\pi f_{\rm FSR})^2}{\gamma^2}\right]\\
+S_{hh}^{\rm SRC}+S_{hh}^{\rm arm}\,,
\end{split}
\end{equation}
where $e^{-2r}=0.1$ represents 10\,dB of observed squeezing. In the design of the FSR detector, we adopt $T_{\rm SRM}=0.0015$, with 10\,dB squeezing and the assumption of 1000\,ppm SRC loss. The quantum noise amplitude result is a factor of $\sqrt{1.3}$ greater than the SRC loss limit. The total quantum noise is shown in Fig.~\ref{fig:FSRbudget}. The other detector parameters are listed in Table.~\ref{ta:par}.

\section{Classical noises}\label{sec:classical}
\def\js#1{{\textcolor{green}{\bf JS: #1}}}

In terms of fundamental classical noises in a gravitational wave detector, the main contributions at kHz frequencies will be from residual gas phase noise and mirror thermal noises. 

\begin{table}
\caption{Parameters of the detector}\label{ta:par}
\begin{ruledtabular}
\begin{tabular}{ccc}
    Parameters &``HF Detector" & ``FSR Detector"  \\
    \hline\\
    Arm length, $L$ & 20\,km & 34\,km\\
    Wavelength & 1064\,nm & 1550\,nm \\
    Mirror Mass & 100\,kg & 500\,kg\\
    Mirror Radius &0.24\,m  &0.34\,m\\
    Mirror thickness &0.25\,m &0.6\,m \\
    ITM/ETM RoC & 14000\,m  &28900\,m\\
    Arm circulating power,\,$ P_{\rm arm}$ &4\,MW &10\,MW \\
    ITM transmittivity & 0.06  & 0.06\\
    SRM transmittivity & 0.1& 0.0015\\
    SRC length &100\,m &50\,m\\
    SRC loss & 1000\,ppm &1000\,ppm\\
    Arm cavity loss & 100\,ppm & 100\,ppm\\
    Squeezing level & 10\,dB & 10\,dB
\end{tabular}
\end{ruledtabular}
\end{table}

\subsection{Residual gas noise}
Even though we operate in high vacuum, there will still be residual gas pressure. The statistical variations of multiple molecular species disturb the laser's phase when they pass across the beam\,\cite{PhysRevA.95.043831}. To calculate this noise, we take the vacuum requirement designed for the Einstein Telescope and specify the residual gas contents, $\rm H_20, H_2, N_2$, and  $\rm O_2 $, with partial pressures of $2.5\times10^{-9}, 5\times10^{-9}, 2.5\times 10^{-10}$ and $10^{-20}$ Pa\,\cite{ET-D,ET2020,Hild_2011} respectively, for both the HF and FSR detectors. Our strategy resembles very closely the implementation in the noise modelling tool GWINC (Gravitational Wave Interferometer Noise Calculator)\,\cite{gitlab}, which uses the approach of Ref.\,\cite{P940008}. We simply reinstated the numerical integration of Equation.~(1) in Ref.~\cite{P940008} rather than maintaining the first order approximation, which strongly diverges above 1\,kHz. Our treatment does pose a greater computational burden but is necessary to accurately model the residual gas noise.

\subsection{Mirror thermal noise }
The mirror thermal noise consists of coating and substrate brownian thermal noise\,\cite{GORODETSKY20086813}, which, according to the fluctuation-dissipation theorem\,\cite{PhysRev.83.34,PhysRev.83.1231,PhysRev.86.702}, come from the thermal fluctuations resulting from the mechanical dissipation in the substrate and the multiple layers of high and low index coatings\,\cite{PhysRevD.91.042002,PhysRevD.95.022001}. They are proportional to the temperature and inversely proportional to the beam size on the test masses. 

In the HF detector, the sensitivity above 1kHz is only limited by the quantum shot noise. This allows us to use the existing Advanced LIGO technology. Thus the HF detector is designed to be operated at room temperature, with a 1064\,nm laser, fused silica substrates and $\rm Ta_2O_5$/$\rm SiO_2$ coatings. Since the thermal noise is not limiting, a relatively smaller beam size allows smaller and lighter test masses. In our case, we choose a 100\,kg mirror with mirror radius of curve (RoC), radius and thickness as shown in Table.~\ref{ta:par}, which are determined by satisfying a beam clipping loss of $\sim$ 1\,ppm (The ratio of mirror size and beam size is $\sim$ 2.63).  
In the FSR detector, the thermal noise will limit the sensitivity. We design the FSR detector at a temperature of 120\,K, with a 1550\,nm laser, silicon substrates and amorphous silicon ($\rm \alpha$-Si) /silicon dioxide ($\rm SiO_2$) optical coatings. It is worthwhile to discuss the calculation of substrate Brownian noise. Via the fluctuation-dissipation theorem, the substrate Brownian noise is related to the mechanical conductance of the substrate itself. As an extended body, the mirror's mechanical conductance is affected by its vibrational eigenmodes, which affect the shape of the substrate surface that the laser impinges upon. 
The vibrational eigenfrequencies lie in the vicity of a few kHz and the lowest frequency is inversely proportional to the mirror size. In the current operating and proposed detectors, the detector's sensitivity in the kHz range is limited by quantum shot noise. In such a case it is common to employ the approach of Ref.~\cite{franc2009mirror,doi:10.1121/1.410467, PhysRevD.62.122002}.
However, this is not the case for the FSR detector, since the quantum noise in the FSR detector is around or below $ 10^{-25}/\sqrt{\rm Hz}$ around 4\,-\,5 kHz and the thermal noise around the substrate resonances is not negligible. It is thus imperative that we model the contribution of each vibrational eigenmode individually, as we are operating in a regime where the low frequency approximation breaks down. This is further exacerbated by us increasing the mirror size in the long baseline detector, which lowers the resonant frequencies yet further for a given fixed aspect ratio. Thus, to minimise the substrate thermal noise, we focus primarily on shifting the resonances to higher frequencies. We achieve this in part by our choice of aspect ratio, which we empirically determine by a finite element method (FEM) to approximately maximise the frequency of the first resonant mode. Consequently, we adopt an aspect ratio of 1.75 in our work. Further optimisation can be achieved by minimising the mirror size (to raise the resonant frequencies) while guaranteeing beam clipping loss of no more than 1\,ppm. The adopted mirror geometry parameters are listed in Table.~\ref{ta:par}. With our optimised mirror design for the FSR detector, we shift all frequencies to be above 6\,kHz. 

To accurately determine the substrate Brownian noise, we require a method that models the noise with reference to the individual contributions of each vibrational eigenmode. Therefore, we implement the approach of \cite{PhysRevD.52.577}, utilising our FEM to determine the mode frequencies and the corresponding surface deformation. This technique is known to be more computationally intensive and, with the first 100 modes considered, we can achieve convergence to approximately 80\% of the noise amplitude given by the direct approach at low frequencies. However, a higher level of convergence is unnecessary as we are interested in the region around the first resonance where the tails of the first few resonant peaks begin to dominate the substrate thermal noise above the combined contributions of all subsequent modes. For a sufficient modelling of the noise spectrum in the frequency band of interest, only the first few modes are necessary.

\section{Summary}

In the era of multi-messenger gravitational-wave astronomy, we now have unprecedented opportunities to better understand 
 the strong field gravitational dynamics and complicated astrophysical processes of a binary neutron star merger, with gravitational wave observations
and  a slew of electromagnetic transients that subsequently arise from the post-merger remnant. The nature  of the post-merger remnant and the lifetime it survives are the key factors for answering the existing open questions, such as the engine of short gamma-ray bursts. 
Previous works have discussed possible ways to probe the post-merger gravitational wave emission with targeted designs of detectors \cite{PhysRevD.99.102004,miao2018towards,ackley2020neutron}.
In this work, we explore the scope of a gravitational wave detector to measure the lifetime of a post-merger remnant star, or constrain its timing to collapse to a black hole. We propose two strategies:  (1) an indirect approach: searching for the peak mode of the post merger waveform, and inferring collapse in the absence of a detection after some time; (2) a direct approach: measuring the 20 mode and/or 22 mode components of the collapse waveform.

Two detector configurations, an HF detector and an FSR detector, are modelled for the two approaches. Here the optimal lengths for detecting the peak mode and the 20 mode are 20\,km and 34\,km, respectively. In the HF detector, we take advantage of a coupled arm cavity and SRC resonance to boost the sensitivity at 2-3\,kHz, which is now limited by the SRC loss. In the
FSR detector, our strategy is to make use of the detector sensitivity at the first FSR, which leads to a long baseline design.
To achieve a loss-limited quantum sensitivity at the FSR frequency, we use an optomechanical detector with negative dispersion to cancel the phase of the sideband gained upon reflection from the arm cavity. This converts the SRC into a signal amplifier. The FSR detector is eventually limited by both quantum and classical noise, in particular, residual gas noise and mirror thermal noise. The frequency dependent antenna response is properly calculated, without using the long wavelength approximation, which is not applicable for our long baseline detector. 

With these optimized detector designs, we realise a SNR of 0.5\,-\,8.6 in the indirect approach and a SNR of 0.3\,-\,1.9 for directly detecting the 20 mode of the collapse waveform. In particular, it seems that a SNR of 1.9 in the direct approach is probably not significant enough to claim detection, as compared to the threshold  ${\rm SNR} \sim 5$ commonly used in black hole ringdown studies \cite{yang2017black}. In the current noise budget the sensitivity of the FSR detector is already limited by both quantum losses and several fundamental classical noises, \textit{i.e.} the mirror thermal noises and the residual gas noise which depends on the facility. Further improving the detector sensitivity not only relies on efforts towards the laser power enhancement for shot-noise reduction, but also on new technologies for the classical noises suppression. These are ongoing investigations in the entire community of the gravitational-wave detection.

\section{Acknowledgements}
We would like to thank LVK collaboration and Andreas Freise for fruitful discussions. T. Z., J. S., J. B., D. M. and H. M. acknowledge the support from the Institute for Gravitational Wave Astronomy at University of Birmingham. Y.C. acknowledges the support from the Department of Astronomy at Beijing Normal University. H. M. is supported by UK STFC Ernest Rutherford Fellowship (Grant No. ST/M005844/11).  
W.E. and H.Y. acknowledge support from an NSERC Discovery grant.  This research
was supported in part by Perimeter Institute for Theoretical Physics.  Research
at Perimeter Institute is supported by the Government of Canada through the
Department of Innovation, Science and Economic Development Canada and by the
Province of Ontario through the Ministry of Research, Innovation and Science.
This research was enabled in part by support provided by SciNet
(www.scinethpc.ca) and Compute Canada (www.computecanada.ca).

\bibliography{bibliography}
\end{document}